\documentstyle[eqsecnum,floats,preprint,aps]{revtex}

\tighten

\begin{document}

\newcommand{\vp}{\varphi}
\newcommand{\CP}{${\cal CP}$}
\newcommand{\C}{${\cal C}$}
\newcommand{\Parity}{${\cal P}$}
\newcommand{\B}{${\cal B}$}
\newcommand{\be}{\begin{equation}}
\newcommand{\ee}{\end{equation}}
\newcommand{\bea}{\begin{eqnarray}}
\newcommand{\eea}{\end{eqnarray}}

\title{Microphysics of Gauge Vortices and Baryogenesis\footnote{Invited talk
at ``Solitons'', Queens University, Kingston, Ontario, Canada, 
July 20-25 1997. To appear in the proceedings.}}

\author{Mark Trodden\footnote{trodden@theory1.phys.cwru.edu.}}

\address{~\\Particle Astrophysics Theory Group \\
Department of Physics \\
Case Western Reserve University \\
10900 Euclid Avenue \\
Cleveland, OH 44106-7079, USA.}

\address{~\\Center for Theoretical Physics \\
Massachusetts Institute of Technology \\
Cambridge, MA 02139, USA}

\maketitle
\begin{abstract}
I give a brief overview of a novel mechanism for generating the baryon 
asymmetry of the universe at the electroweak scale. This scenario 
circumvents the need for a strongly first order electroweak phase transition 
by utilizing gauged topological solitons to realize the requisite departure 
from thermal equilibrium.
\end{abstract}

\setcounter{page}{0}
\thispagestyle{empty}
\vfill
\baselineskip 14pt

\noindent CWRU-P12-97

\noindent MIT-CTP-2665 

\eject

\vfill

\eject

\baselineskip 24pt plus 2pt minus 2pt

\section{Introduction}
To a very good approximation, modern particle physics theories treat
particles and their antiparticles in a symmetric manner. In particular this 
means that the physics describing the production, interactions and
destruction of antimatter is the same as that describing the corresponding 
processes for
matter. This fact is in stark contrast to terrestrial and cosmological
observations which reveal that the universe appears to consist almost
entirely of matter, with negligible antimatter on all scales up to the
causal (Hubble) distance\cite{CRG 97}. 

At a quantitative level, the outstanding success of the theory of 
primordial nucleosynthesis requires that 

\be
2\times 10^{-10} < \eta \equiv \frac{n_b - n_{\bar b}}{s} < 7
\times 10^{-10} \ ,
\ee
where $n_b$ denotes the number density of baryons, $n_{\bar b}$ that of
antibaryons and $s$ denotes the entropy density.

Electroweak baryogenesis is an attempt to understand how this number could 
have been dynamically generated at the electroweak scale, 
$\eta_{EW}\sim 10^2$GeV, 
from baryon number symmetric initial conditions in the early 
universe (for a review and references see\cite{MSreview}.
Here I present a brief overview of a variant of such a scenario in which
a pivotal role is played by gauged topological 
solitons\cite{{DMEWBG},{Models}}.

\section{The Electroweak Theory and Sakharov}
Sakharov\cite{Sakharov} identified the three necessary conditions for a 
particle physics model to produce a net baryon asymmetry. These are a 
violation of baryon number (\B) conservation, violations of charge 
conjugation (\C) and charge-parity (\CP) symmetries and a departure from 
thermal equilibrium. The electroweak theory includes all three 
ingredients. 
Baryon number is only violated nonperturbatively at the quantum level as a 
consequence of the anomaly\cite{tHooft}. The relevant process at zero 
temperature is an exponentially suppressed tunnelling under the energy 
barrier which separates inequivalent vacua in gauge and Higgs field 
configuration space. However, at temperatures around or above the critical 
temperature of the electroweak phase transition (EWPT) thermal effects cause 
classical transitions over the barrier\cite{KRS,AM}. 

The standard electroweak
theory is maximally \C\ violating due to the V-A nature of the interactions.
The only violation of \CP\ in the electroweak sector of the standard model 
occurs in the CKM matrix.
Electroweak baryogenesis scenarios using this source of \CP\ violation are
generally expected to produce a BAU which is far too small and so
it is usual to consider extending the standard model to include new
sources of \CP\ violation. Examples of extensions are the two-Higgs doublet 
model with explicit renormalizable \CP\ violating 
terms and effective theories of the standard model plus nonrenormalizable 
operators, some of which are \CP\ odd.

Finally, the departure from thermal equilibrium is typically achieved by 
assuming
that the electroweak phase transition is strongly first order so that 
the violent conversion of the symmetric phase to 
the broken phase results in non-equilibrium conditions near the bubble 
walls separating the phases. 
In the minimal electroweak theory this assumption now appears unlikely to 
be true\cite{KLRS} and here I present an alternative realization of the 
third Sakharov condition.

\section{Electroweak Symmetry Restoration around Vortices}
The contents of this section can be extended to monopoles and domain walls.
Here I restrict the discussion to cosmic strings and to a particular
example\cite{P&D 93}. Consider a local $U(1)$ cosmic string, formed at a 
scale 
$\eta>\eta_{EW}$, that couples to the electroweak model. Further, assume 
that the gauge fields corresponding to this higher symmetry scale acquire an 
extra mass at the electroweak scale.
Let the string's gauge field be $R_{\mu}$. The
coupling between the string and the electroweak sector is through the 
covariant derivative
\be
D_{\mu}\Phi=\left(\partial_{\mu}- \frac{1}{2}ig{\bf \tau}.{\bf W}_{\mu} -
\frac{1}{2}ig'B_{\mu}-\frac{1}{2}ig''R_{\mu}\right)\Phi \ .
\ee
where $B_{\mu}$ is the hypercharge field and $W^a_{\mu}$ are the weak
isospin fields. Since $\eta>\eta_{EW}$, we treat $R_{\mu}$ as a 
Nielsen-Olesen background. It can be shown the minimal energy is achieved 
when the electroweak symmetry is restored around the defect out to a radius 

\be
R_s \sim \eta_{EW}^{-1}\ ,
\ee
up to couplings. A similar effect occurs around superconducting cosmic 
strings\cite{P&D 93,MT 94}. Within this symmetric electroweak region, \B\ 
violating processes should not be exponentially suppressed.

\section{Defect-Mediated Electroweak Baryogenesis}
The central idea is to compare the motion of the phase interfaces of an
evolving network of cosmic strings with bubble walls at a first order phase
transition. Denote the baryon to entropy ratio produced by
a bubble wall scenario by $n_b^{(0)}/s$. This is generated when a wall passes
points in space and false vacuum is converted to true. To compute the 
baryon asymmetry produced by strings, there are two principal effects to be 
taken into account.

First, there is an effect due to the geometry of the defects. Bubble walls
sweep out the whole of space whereas a network of cosmic strings only sweeps
out a volume $V_{BG}$ in one Hubble time after the EWPT. This leads to a 
suppression factor 
$\delta_1 \equiv V_{BG}/V$, where $V$ is the total volume, compared to bubble 
wall baryogenesis. It turns out that only strings formed close to the 
electroweak scale yield a measurable asymmetry. In this case ($\eta$ close
to $\eta_{EW}$), the network is still in the friction-dominated epoch at the
EWPT. For simplicity we focus on the contribution of long strings and assume 
one per correlation volume at the formation temperature, $\eta$. Since the
correlation length $\xi$ obeys\cite{KEH}
$\xi(\eta)\sim \eta^{-1}$ and $\xi(T) \sim \xi(\eta)(\eta/T)^{5/2}$, we 
obtain

\be
\delta_1 \sim v\left(\frac{\eta_{EW}}{\eta}\right)^{3/2} \ ,
\ee
where $v$ is the string velocity.

The second effect is due to cancellations between competing processes at the 
two faces of the
string. The trailing edge of the defect behaves analogously to a bubble wall.
In contrast, at the leading edge of the defect true vacuum is converted to 
false and in such a process \CP\ violation works in the opposite way. Thus,
at the leading edge antibaryons are produced. However, cancellation is not 
complete. At the trailing edge the baryons remain ``frozen" into the broken 
phase whereas at the leading edge the
antibaryons spend the core passage time $\tau=R_s/v$ in the symmetric
phase where they can equilibrate to lower \B\ through anomalous processes.
This leads to a further suppression factor

\be
\delta_2 \sim 1-\exp(-\Gamma \tau) \sim \Gamma\tau \ ,
\ee
where $\Gamma$ is the rate of \B\ violating events.

Thus, the final baryon to entropy ratio produced by a network of cosmic
strings produced not too far above the electroweak scale is

\be
\frac{n_b}{s} = \frac{n_b^{(0)}}{s}\delta_1\delta_2 \ .
\ee

Since bubble wall calculations can give $n_b^{(0)}/s \sim 10^{-6} - 10^{-8}$,
choosing $\eta \sim 1$TeV can yield a number consistent with the 
requirements of nucleosynthesis. It is interesting to note that particle
physics models giving rise to such strings (and other solitons) at the TeV 
scale exist in a number of places in the literature\cite{Models}.

\section{Conclusions}
If the electroweak phase transition is not strongly first order, traditional
scenarios for electroweak baryogenesis seem unlikely to succeed. In that 
case, an
alternative realization of the third Sakharov condition is necessary. 
Gauged topological solitons, in particular cosmic strings, can provide
such a realization due to the restoration of the electroweak symmetry
around their cores. The final baryon to entropy ratio generated in such
models is suppressed relative to bubble scenarios but for strings formed
at the TeV scale and optimistic parameter choices an acceptable asymmetry
can still result. The group structure necessary to produce such TeV
defects is present in some popular particle physics models and, if uncovered 
in 
accelerator experiments, would imply the cosmological existence of the
required gauge solitons.

\acknowledgments

I would like to thank the conference organisers for putting together such an
enjoyable meeting and in particular for providing ample time for 
discussion between the sessions. I also owe thanks to my collaborators in
the work described here, Robert Brandenberger, Anne-Christine Davis and 
Tomislav Prokopec. This work was supported by the Department of Energy 
(D.O.E.), the National Science Foundation (N.S.F.) and by funds provided
by Case Western Reserve University.

\end{document}